\documentclass{ws-ijmpa-blank-new}
\usepackage[super,compress]{cite}
\usepackage{graphicx}

\usepackage{amssymb,epsf,latexsym,amssymb,amsmath,mathrsfs}
\usepackage[english]{babel}
\usepackage{dsfont}  %%needed for \openoneFRK

\newcommand{\beq}{\begin{equation}}
\newcommand{\eeq}{\end{equation}}
\newcommand{\beqa}{\begin{eqnarray}}
\newcommand{\eeqa}{\end{eqnarray}}
\newcommand{\bsubeqs}{\begin{subequations}}
\newcommand{\esubeqs}{\end{subequations}}

\newcommand{\openoneFRK}{\mathds{1}}  %%retex4 has \openone


\begin{document}
\markboth{F.R. Klinkhamer and J.M. Queiruga}
{A stealth defect of spacetime}

\catchline{}{}{}{}{}
%
%%%%%%%%%%%%%%%%%%%%%%%%%%%%%%%%%%%%%%%%%%%%%%%%%%%%%%%%%%%%%%%%%%%%

\title{\vspace*{-11mm}
A stealth defect of spacetime}

\author{F.R. Klinkhamer}

\address{Institute for Theoretical Physics, Karlsruhe Institute of
Technology (KIT),\\ 76128 Karlsruhe, Germany\\
frans.klinkhamer@kit.edu}

\author{J.M. Queiruga}
\address{\mbox{Institute for Theoretical Physics,
Karlsruhe Institute of Technology (KIT),}\\
76128 Karlsruhe, Germany\\
and\\
\mbox{Institute for Nuclear Physics,
Karlsruhe Institute of Technology  (KIT),}\\
\mbox{Hermann-von-Helmholtz-Platz 1,
76344 Eggenstein-Leopoldshafen, Germany}\\
jose.queiruga@kit.edu}

\maketitle

%\begin{history}  %%AUTHOR ADDITION: removed for preprint lay-out
%\received{Day Month Year}
%\revised{Day Month Year}
%\end{history}

\begin{abstract}
We discuss a special type of Skyrmion spacetime-defect solution,
which has a positive energy density of the matter fields
but a vanishing asymptotic gravitational mass.
With a mass term for the matter field added to the action
(corresponding to massive
``pions'' in the Skyrme model), this particular
soliton-type solution has no long-range fields
and can appropriately be called a ``stealth defect.''
\end{abstract}
\vspace*{.0\baselineskip}
{\footnotesize
\vspace*{.25\baselineskip}
\noindent \hspace*{5mm}
\emph{Journal}: Mod. Phys. Lett. A \textbf{33} (2018) 1850127
\vspace*{.25\baselineskip}
\newline
\hspace*{5mm}
\emph{Preprint}: arXiv:1805.04091  %%KA-TP-13-2018
}
\vspace*{-5mm}\newline
\ccode{PACS Nos.: 04.20.Cv, 04.20.Gz}

%\tableofcontents

\section{Introduction}
\label{sec:Introduction}

An explicit \emph{Ansatz} for a Skyrmion spacetime
defect has been presented several years ago~\cite{Klinkhamer2014-prd}.
It is now clear that this soliton-type defect solution can have positive
gravitational mass but also negative gravitational
mass if the defect length scale is small enough~\cite{KlinkhamerQueiruga2018-prd}
(additional numerical results can be found in Ref.~\citen{Guenther2017}).
As the gravitational mass of such a spacetime-defect solution
is a continuous variable, there must also be
special spacetime defects with vanishing gravitational
mass. These defects with positive energy density of
the matter fields and zero asymptotic gravitational
mass will be called ``stealth defects.''
In this article, we will describe the stealth defects
in more detail.

Before we turn to the detailed discussion
of stealth defects, it may be helpful to recall the
main properties of the negative-gravitational-mass
Skyrmion spacetime defect. In fact, the origin of the
negative gravitational mass of this type of spacetime defect lies
in the nontrivial topological structure of spacetime itself.
The presence of the spacetime defect, obtained by surgery from
Minkowski spacetime, allows for a degenerate metric at the defect
surface. As explained in
Refs.~\citen{KlinkhamerQueiruga2018-prd,Guenther2017},
this degeneracy makes the
positive-mass theorems not directly applicable and, at the
same time, removes possible​ curvature singularities.

In Ref.~\citen{KlinkhamerQueiruga2018-prd},
we already noted that a special choice of defect scale,
with appropriate boundary conditions,
results in a vanishing asymptotic gravitational mass.
For such a special defect scale,
there exists no globally regular solution with the
standard (zero-effective-mass) boundary conditions at the defect surface.
Therefore, our globally regular solution with nonstandard (negative-effective-mass) boundary conditions at the defect surface
may be considered to be
a natural consequence of the field equations in this setup.
Furthermore,  the combination of nontrivial spacetime topology
and nontrivial target-space topology appears to allow for the existence
of stable static solutions due to topological stabilization. This
topological stability results
from the use of an $SO(3)$ Skyrme field that matches the nontrivial
space topology [basically $\mathbb{R}P^3 \sim SO(3)$;
see Ref.~\citen{Klinkhamer2014-mpla} for a review].​​  

The outline of the present article is as follows.
In Sec.~\ref{sec:Setup}, we give the necessary background for
the Skyrmion spacetime-defect solution and define its gravitational
mass.
In Sec.~\ref{sec:Zero-gravitational-mass-defect}, we obtain
a vanishing gravitational mass of a particular
Skyrmion spacetime-defect solution, primarily by numerical
methods but, for a limiting case, also analytically.
In Sec.~\ref{sec:Discussion}, we give a brief discussion
of how the stealth defect may manifest itself or, rather,
stays hidden for most of the time.
In Sec.~\ref{sec:Outlook},
we put our stealth-defect solution in a larger context,
and compare the stealth defect with a so-called invisibility cloak.

%%\newpage%%tmp
\section{Setup}
\label{sec:Setup}

\subsection{Spacetime manifold}
\label{subsec:Spacetime-manifold}

The spacetime manifold considered has been
discussed extensively in Refs.~\citen{Klinkhamer2014-prd,Klinkhamer2014-mpla},
so that we can be brief.

\begin{figure*}[t]  %% fig01-v1.eps=fig01-v2.eps
\begin{center} %%\end{center}
\includegraphics[width=0.5\textwidth]{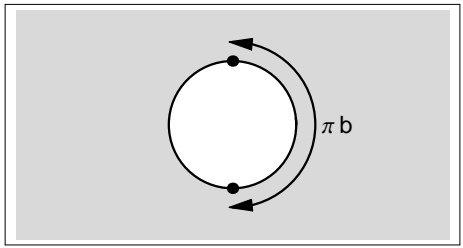}
%{skyrmion-spacetime-defect_fig-sketch_v5.eps}
\end{center}
\caption{Three-space $M_3$ obtained by surgery on
three-dimensional Euclidean space $E_3$:
the interior of a ball with radius $b$ is removed
and antipodal points on the boundary of the ball are
identified (as indicated by the dots).
The ``long distance'' between antipodal points
\mbox{equals $\pi\, b$ in the original three-space $E_3$.}  
(Figure taken from Ref.~\citen{Klinkhamer2014-prd}.)}
\label{fig01}
\end{figure*}

The four-dimensional spacetime manifold has the following
topology:
\bsubeqs\label{eq:M4-M3-topology}
\beqa\label{eq:M4}
M_4 &=& \mathbb{R} \times M_3\,,
\\[2mm]
\label{eq:M3}
M_3 &=&
\mathbb{R}P^3 - p_\infty\,,
\eeqa
\esubeqs
where $p_\infty$ corresponds to the ``point at spatial infinity.''
The three-space $M_3$ is a noncompact, orientable, nonsimply-connected
manifold without boundary. A sketch of the space manifold $M_3$
is given in Fig.~\ref{fig01}.

A particular covering of $M_3$ has three charts of coordinates,
labeled by $n=1,2,3$, and details can be found in
Ref.~\citen{Klinkhamer2014-mpla}.
These coordinates resemble the standard spherical coordinates and
each coordinate chart surrounds one of the three Cartesian coordinate axes
but does not intersect the other two axes. The coordinates are denoted
\beq\label{eq:XnYnZn}
(X_n,\,  Y_n,\, Z_n)\,,
\eeq
for  $n=1,\,2,\,3$. In each chart,
there is one polar-type angular coordinate of finite range,
one azimuthal-type angular coordinate of finite range, and
one radial-type coordinate with infinite range.
Specifically, the coordinates have the following ranges:%
\bsubeqs\label{eq:XnYnZn-ranges}    
\begin{align} %%uses package amsmath, and \begin{align*} for without eq. nos.
\label{eq:X1Y1Z1-ranges}
X_{1} &\in (-\infty,\,\infty) \,,
&
Y_{1} &\in (0,\,\pi)\,,
&
Z_{1} &\in (0,\,\pi)\,,
\\[2mm]
\label{eq:X2Y2Z2-ranges}
X_{2} &\in (0,\,\pi)\,,
&
Y_{2} &\in (-\infty,\,\infty)\,,
&
Z_{2} &\in (0,\,\pi)\,,
\\[2mm]
\label{eq:X3Y3Z3-ranges}
X_{3} &\in  (0,\,\pi)\,,
&
Y_{3} &\in  (0,\,\pi)\,,
&
Z_{3} &\in  (-\infty,\,\infty)\,.
\end{align} 
\esubeqs
The different charts overlap in certain regions, see
Ref.~\citen{Klinkhamer2014-mpla} for further details.
In the present article, we
focus on the $n=2$ chart
and drop the suffix $2$ on the coordinates,
\begin{align} %%uses package amsmath, and \begin{align*} for without eq. nos.
\label{eq:XYZ-ranges}
X &\in (0,\,\pi)\,,
&
Y &\in (-\infty,\,\infty)\,,
&
Z &\in (0,\,\pi)\,,
\end{align} 
where $Y=0$ indicates the position of the defect surface.
Together with the time coordinate $T$, we collectively
denote these four spacetime coordinates as $X$.
Later, the metric signature will be taken as $(-+++)$.

%%\newpage%%tmp
\subsection{Action}
\label{subsec:Action}

The matter fields and their interactions have also been described
in Ref.~\citen{Klinkhamer2014-prd}, but, as we will make
an addition to the theory, we will give some details.

The action reads ($c=\hbar=1$)
\bsubeqs\label{eq:action-Lgrav-Lmat-omegamu}
\beqa\label{eq:action}
\hspace*{-6mm}
S &=&
 \int_{M_4} d^4X\,\sqrt{-g}\,
\Big(\mathcal{L}_{\text{grav}}+ \mathcal{L}_{\text{mat}}\Big)\,,
\\[2mm]
\label{eq:Lgrav}
\hspace*{-6mm}
\mathcal{L}_{\text{grav}}&=& \frac{1}{16\pi G_{N}}\:R\,,
\\[2mm]
\hspace*{-6mm}
\label{eq:Lmat}
\mathcal{L}_{\text{mat}}&=&
\frac{f^2}{4}\:\text{tr}\Big(\omega_\mu\,\omega^\mu\Big)
+\frac{1}{16\, e^2}\: \text{tr}\Big(\left[\omega_\mu,\,\omega_\nu\right]
\left[\omega^\mu,\,\omega^\nu\right]\Big)
 +	\frac{1}{2}\,m^2\, f^2\,\text{tr}\Big(\Omega-\openoneFRK_{3}\Big)\,,
\\[2mm]
\label{eq:omegamu}
\hspace*{-6mm}
\omega_\mu &\equiv& \Omega^{-1}\,\partial_\mu\,\Omega\,,
\eeqa
\esubeqs
with a Skyrme-type scalar field $\Omega(X)\in SO(3)$.
New in \eqref{eq:Lmat} is the mass term proportional to $m^2$.
Indeed, consider the ``pions'' $\pi^a$ defined by the following
expansion:
\beq
\label{eq:Omega-pi-a}
\Omega(X)=\exp\Big[S^a\, \pi^a(X)/f \,\Big]\,,
\eeq
with an implicit sum over $a=1,\,2,\,3$,
and $3\times 3$ matrices $S^a$ given by
\beqa
\label{eq:S123-def}
S_1 &\equiv&  \left(
                \begin{array}{ccc}
                  0     & 0  &   0 \\
                  0     & 0  &   1 \\
                  \;0\; & -1 & \;0\; \\
                \end{array}
              \right)\,,
\quad
S_2 \equiv  \left(
                \begin{array}{ccc}
                  \;0\; & \;0\; & -1 \\
                    0   &   0   & 0 \\
                    1   &   0   & 0 \\
                \end{array}
              \right)\,,
\quad
S_3 \equiv  \left(
                \begin{array}{ccc}
                  0  &   1   &   0 \\
                  -1 & \;0\; & \;0\; \\
                  0  &   0   &   0 \\
                \end{array}
              \right)\,.
\eeqa
Incidentally, a useful discussion of the $SO(3)$ Lie group
can be found in App.~C, pp. 436--438 of Ref.~\citen{DeWitSmith1986}.
Expanding the $\Omega$ field in \eqref{eq:Lmat} 
with the Taylor series from \eqref{eq:Omega-pi-a}   
and using $\text{tr}(S^a S^b)=-2\,\delta^{ab}$, we have
\beq
\mathcal{L}_{\text{mat}}=
-\frac{1}{2}\,\partial_\mu \pi^a\partial^\mu \pi^a
-\frac{1}{2}\,m^2\,\pi^a\pi^a+ \cdots \,,
\eeq
which corresponds to three real scalars with equal mass $m$.

The parameters of the theory
\eqref{eq:action-Lgrav-Lmat-omegamu} are
Newton's gravitational coupling constant $G_{N}$
and the energy scale $f>0$ from the kinetic matter term in the
action, together with the mass $m$ which we assume
to be of order $f$.
The first two parameters can be combined in the following
dimensionless parameter:
\begin{eqnarray}\label{eq:dimensionless-eta}
\widetilde{\eta}&\equiv& 8\pi\, G_{N}\, f^2 \geq 0\,.
\end{eqnarray}
The quartic Skyrme term in \eqref{eq:Lmat} has an additional
dimensionless real coupling constant,
\begin{eqnarray}\label{eq:dimensionless-Skyrme-e}
e &\in&  (0,\,\infty)\,.
\end{eqnarray}

%%\newpage%%tmp
\subsection{Spacetime-defect solution}
\label{subsec:Spacetime-defect-solution}

The self-consistent \emph{Ans\"{a}tze} for the metric
and the $SO(3)$ matter field have been presented in
Ref.~\citen{Klinkhamer2014-prd}.
For completeness, we give the relevant expressions:
\bsubeqs\label{eq:Ansaetze}
\beqa\label{eq:metric-Ansatz}
\hspace*{-8mm}
ds^2 &=&
 -\big[\mu(W)\big]^2\, dT^2
 +\big(1-b^2/W\big)\,\big[\sigma(W)\big]^2\,(d Y)^2
\nonumber\\[1mm]
&&
+W \Big[(d Z)^2+\sin^2 Z\, (d X)^2 \Big]\,,
\\[2mm]
\label{eq:hedgehog-Ansatz}
\hspace*{-8mm}
\Omega(X) &=&
\cos\big[F(W)\big]\;\openoneFRK_{3}
-\sin\big[F(W)\big]\;
\widehat{x}\cdot \vec{S}
%%\nonumber\\&&
+\big(1-\cos\big[F(W)\big]\big)\;
\widehat{x} \otimes \widehat{x}\,,
\\[2mm]
\label{eq:W-definition}
\hspace*{-8mm}
W &\equiv& b^2+Y^2\,,
\eeqa
\esubeqs
with the matrices $S^a$ from \eqref{eq:S123-def}
and the unit 3-vector $\widehat{x} \equiv \vec{x}/|\vec{x}|$
from the Cartesian coordinates $\vec{x}$ defined in terms
of the coordinates $X$, $Y$, and $Z$ (details are in
Ref.~\citen{Klinkhamer2014-mpla}).

The fields of these \emph{Ans\"{a}tze} are characterized
by the defect length scale $b>0$  \mbox{(cf. Fig.~\ref{fig01}).}
Henceforth, we will use the \emph{Ansatz} functions
$F(w)$, $\sigma(w)$, and $\mu(w)$ defined
in terms of the following dimensionless variable:
\beq\label{eq:w}
w \equiv (e\,f)^2\;W \in \big[(y_0)^2,\,\infty\big)\,,
\eeq
where $y_0$ is the dimensionless defect scale,
\beq\label{eq:y0}
y_0 \equiv e\,f\;b \in  \big(0,\,\infty\big)\,.
\eeq

The \emph{Ans\"{a}tze} \eqref{eq:Ansaetze} reduce the field equations 
from \eqref{eq:action-Lgrav-Lmat-omegamu}
to the following three ordinary differential equations (ODEs):
\bsubeqs\label{eq:ODEs-with-pion-mass}
\begin{eqnarray}
\label{eq:ODE-sigma-with-pion-mass}
\hspace*{-10mm}
4\,w\,\sigma'(w)
&=&
+\sigma(w)\,\left[
\left[1-\sigma^2(w)\right]
+\widetilde{\eta}\, \frac{2}{w}\,
\Big(A(w)\,\sigma^2(w) + C(w)\,\left[w\,F'(w)\right]^2\Big)
\right]\nonumber\\
&&
+2\, w\,\widetilde{m}^2\,\widetilde{\eta}\,
\sigma^3(w)\,
%%frk: have now corrected calculation for fig.04, no visible change
\sin^2\frac{F(w)}{2}\,,
\\[2mm]
%\end{eqnarray}
%\begin{eqnarray}
\label{eq:ODE-mu-with-pion-mass}
\hspace*{-10mm}
4\,w\,\mu'(w)
&=&
-\mu(w)\,\left[
\left[1-\sigma^2(w)\right]
+ \widetilde{\eta}\, \frac{2}{w}\,
\Big(A(w)\,\sigma^2(w) - C(w)\,\left[w\,F'(w)\right]^2\Big)
\right]\nonumber\\
&&
-2\, w\,\widetilde{m}^2\,\widetilde{\eta}\,\sigma^2(w)\,\sin^2\frac{F(w)}{2}\,,
\\[2mm]
%\end{eqnarray}
%\begin{eqnarray}
\label{eq:ODE-F-with-pion-mass}
\hspace*{-0mm}
C(w)\,w^2\,F''(w)
&=&
+\sigma^2(w)\,\sin F(w)
\,\left( \sin^2\frac{F(w)}{2}+\frac{w}{2}  \right)
%%\nonumber\\&&
-\frac{1}{2}\,C(w)\,\sigma^2(w)\,w\,F'(w)
\nonumber\\&&
\times
\,\left[1-4\,\widetilde{\eta}\,\frac{1}{w}\, \sin^2\frac{F(w)}{2}
 \,\left(\sin^2\frac{F(w)}{2} +w  \right) \right]
\nonumber\\&&
-w\,F'(w)
\,\Big[w\,F'(w)\,\sin F(w)+w \Big]
%%\nonumber\\&&
+\frac{\widetilde{m}^2}{2}\,w^2\,\sigma^2(w)\,\sin\frac{F(w)}{2}
\nonumber\\&&
\times
\left[\cos\frac{F(w)}{2}
+2\,\widetilde{\eta}\,C(w)\,\sin\frac{F(w)}{2}F'(w)\right]\,,
\end{eqnarray}
\esubeqs
with a further dimensionless quantity
\beq\label{eq:mtildesquare}
\widetilde{m}^2\equiv \frac{1}{e^2\, f^2}\,m^2 \,,
\eeq
and the following auxiliary functions:
\bsubeqs\label{eq:A-C-def}
\beqa
\label{eq:A-def}
A(w) &\equiv& 2\,\sin^2\frac{F(w)}{2}
\left(\sin^2\frac{F(w)}{2}+w\right)  \,,
\\[2mm]
%\end{eqnarray}
%\begin{eqnarray}
\label{eq:C-def}
C(w)
&\equiv&
4\,\sin^2\frac{F(w)}{2}+w \,.
\eeqa
\esubeqs
These ODEs are to be solved with the following boundary conditions:
\bsubeqs\label{eq:bcs-F-sigma-mu}
\beqa
\label{eq:bcs-F}
F(b^2)    &=& \pi\,,\quad F(\infty) = 0 \,,
\\[2mm]
\label{eq:bcs-sigma}
\sigma(b^2) &\in&  (0,\,\infty) \,,
\\[2mm]
\label{eq:bcs-mu}
\mu(b^2)&\in&  (0,\,\infty) \,,
\eeqa\esubeqs
where the zero values for $\sigma(b^2)$ and $\mu(b^2)$
have been excluded,
so that the reduced field equations are well-defined
at $Y=0$ (see Sec.~3.3.1 of Ref.~\citen{Guenther2017}).

The solutions of the ODEs \eqref{eq:ODEs-with-pion-mass}
with boundary conditions \eqref{eq:bcs-F-sigma-mu}
behave as follows asymptotically ($w\to\infty$):
\bsubeqs\label{eq:F-sigma-mu-asymptotic}
\beqa\label{eq:F-asymptotic}
F(w) &\sim& \widetilde{k}\,\exp\Big[-\widetilde{m}\,\sqrt{w}\,\Big]\,
            \frac{1+\widetilde{m}\,\sqrt{w}}{w}\,,
\\[2mm]
\label{eq:sigma-asymptotic}
\sigma(w) &\sim&
\sqrt{1\big/\big(1-\widetilde{l}/\sqrt{w}\big)} \,,
\\[2mm]
\label{eq:mu-asymptotic}
\mu(w)&\sim& \sqrt{1-\widetilde{l}/\sqrt{w}} \,,
\eeqa
\esubeqs
with constants $\widetilde{k}$ and $\widetilde{l}$. Observe that
another contribution
$F(w) \propto \exp\big[+\widetilde{m}\,\sqrt{w}\,\big]$
is eliminated by the boundary condition
\mbox{$F(\infty)=0$.}

The reduced expressions for the Ricci curvature scalar $R(w)$,
the Kretschmann curvature  scalar $K(w)$,
and the negative of the 00-component of
the Einstein tensor $E^{\mu}_{\;\;\nu}(w)$
$\equiv$ $R^{\mu}_{\;\;\nu}(w) -(1/2)\,R(w)\,\delta^{\mu}_{\;\;\nu}$
have been given in Appendix~B of Ref.~\citen{Klinkhamer2014-prd}.
Here, we present the reduced expression for the
00-component of the energy-momentum tensor $T^{\mu}_{\,\,\nu}(w)$
which corresponds to the negative of the energy density $\rho(w)$,
\beqa
\label{eq:T-up0-down0}
\hspace*{-10mm}
T^0_{\,\,0}(w)
&=&
-f^2\,(e\,f)^2\, \frac{2}{w^2\,\sigma^2(w)}\,
\nonumber\\
\hspace*{-10mm}
&&
\times
\Bigg(
  A(w)\,\sigma^2(w)
+ C(w)\,\big[w\,F'(w)\big]^2
+ \widetilde{m}^2\,w^2\,\sigma^2(w) \,\sin^2\frac{F(w)}{2}
\Bigg) \,.
\eeqa
The expression in the large brackets on the right-hand side
of \eqref{eq:T-up0-down0} already appears on the right-hand side
of \eqref{eq:ODE-sigma-with-pion-mass}, which results from
the 00-component of the Einstein equation.

%%\newpage%%tmp
\subsection{Defect solution for $\mathbf{\widetilde{\eta}=0}$}
\label{subsec:Defect-solution-for-etatilde-zero}

Consider, next, the special theory with $\widetilde{\eta}=0$.
Then, the ODE \eqref{eq:ODE-sigma-with-pion-mass} for $\sigma(w)$
and the ODE \eqref{eq:ODE-mu-with-pion-mass} for $\mu(w)$
become independent
of the matter function $F(w)$ and can be solved analytically.
We obtain the following Schwarzschild-type solutions
(cf. Sec.~3 of Ref.~\citen{Klinkhamer2014-mpla}):
\bsubeqs\label{eq:Schwarzschild-type-solutions-mu-sigma}
\beqa
\label{eq:Schwarzschild-type-solutions-mu}
\overline{\mu}(w)&=& \sqrt{1-\widehat{l}/\sqrt{w}}\,,
\\[2mm]
\label{eq:Schwarzschild-type-solutions-sigma}
\overline{\sigma}(w) &=& \big[\overline{\mu}(w)\big]^{-1}\,,
\eeqa\esubeqs
where, for globally regular solutions, the
constant $\widehat{l}$ takes values in the following range:
\beq\label{eql-bar}
\widehat{l} \in (-\infty,\, y_0)\,.
\eeq

At $\widetilde{\eta}=0$,
the remaining ODE \eqref{eq:ODE-F-with-pion-mass}
for $F(w)$ is reduced to the following form:
\beqa\label{eq:F-ODE-zero-etatilde}
\hspace*{0mm}
C(w)\,w^2\,F''(w)
&=&
+\overline{\sigma}^2(w)\,\sin F(w)
\,\Bigg[ \sin^2\frac{F(w)}{2}+\frac{w}{2}  \Bigg]
%%\nonumber\\&&
-\frac{1}{2}\,C(w)\,\overline{\sigma}^2(w)\,w\,F'(w)
\nonumber\\
&&
-w\,F'(w)
\,\Bigg[w\,F'(w)\,\sin F(w)+w \Bigg]
\nonumber\\
&&
+\frac{\widetilde{m}^2}{4}\,w^2\,\overline{\sigma}^2(w)\,\sin F(w)\,,
\eeqa
with $\overline{\sigma}(w)$ given by
\eqref{eq:Schwarzschild-type-solutions-sigma} and
$C(w)$ by \eqref{eq:C-def}.
The ODE \eqref{eq:F-ODE-zero-etatilde} with
boundary conditions \eqref{eq:bcs-F} solely
depends on the constant $\widehat{l}$,
but still cannot be solved analytically.

Physically, the special theory with $\widetilde{\eta}=0$
gives a close approximation to the theory with
the matter energy scale $f$ far below the reduced Planck energy
scale (the mass $m$ is assumed to be of the same order as $f$),
\beqa\label{eq:matter-theory-f2bound}
f^2  &\ll&    \big(E_\text{planck}\big)^2
     \equiv   1/(8\pi G_{N})
     \approx  \big(2.44\times 10^{18}\;\text{GeV}\big)^2\,.
\eeqa
As discussed in App.~A of Ref.~\citen{KlinkhamerQueiruga2018-prd},
the theory with a low energy scale \eqref{eq:matter-theory-f2bound}
and a quartic coupling constant $e \leq 1/\xi$ has
anti-gravitating defects if their length scale $b$ is set by
the Planck length,
$b =\xi\, l_\text{planck}$
for a constant $\xi \gtrsim 1$ and with
$l_\text{planck}\equiv \hbar c/E_\text{planck}
\approx 8.10 \times 10^{-35}\;\text{m}$.

%%\newpage%%tmp
\subsection{Gravitational mass of the defect}
\label{subsec:Gravitational-mass-of-the-defect}

Finally, introduce the following dimensionless mass-type variable:
\beq\label{eq:l}
l(w) \equiv \sqrt{w}\,\left[1-\frac{1}{\sigma^{2}(w)}\right]\,,
\eeq
which corresponds to a Schwarzschild-type behavior
of the square of the metric function,
\beq\label{eq:sigmaw-from-lw}
\sigma^{2}(w)=\frac{1}{1-l(w)/\sqrt{w}\,}\,.
\eeq
The Arnowitt--Deser--Misner (ADM)
mass~\cite{ADM1959} is then obtained by the following limit:%
\bsubeqs\label{eq:M-ADM}
\beqa
M_{\text{ADM}} &=&
\frac{l_\infty}{2\, G_{N} \,e\,f}\,,
\\[2mm]
l_\infty &\equiv& \lim_{w\rightarrow\infty} l(w) \,.
\eeqa
\esubeqs
The task, now,  is  to find solutions with vanishing ADM mass.

%%\newpage%%tmp
\section{Zero-gravitational-mass defect}
\label{sec:Zero-gravitational-mass-defect}

\begin{figure*}[ht]   %%[h!]
\vspace*{-0mm}
\begin{center} %%\end{center}
\includegraphics[width=0.925\textwidth]{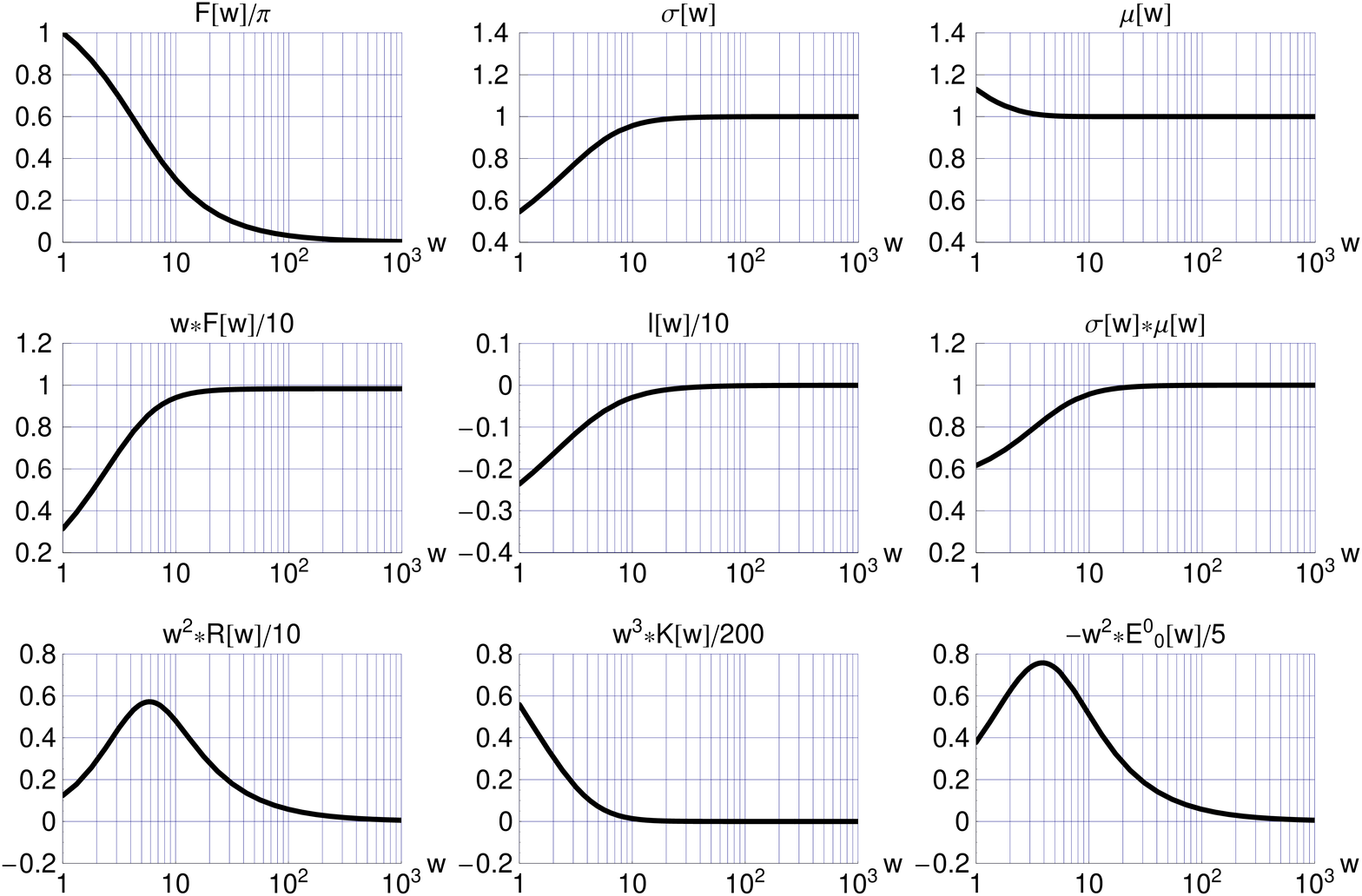}
\end{center}
\vspace*{-0mm}
\caption{(Top row)
\textit{Ansatz} functions $F(w)$, $\sigma(w)$, and $\mu(w)$
of the numerical solution of the ODEs \eqref{eq:ODEs-with-pion-mass}
with $\widetilde{m}=0$. The parameters
are $\widetilde{\eta}\equiv 8\pi\, G_{N}\, f^2  =1/10$ and
$y_0 \equiv ef b =1$.
The boundary conditions at the defect surface $w=y_0^2=1$ are:
$F/\pi=    1.00000   %%1
$,
$F^\prime=-0.569030  %%-0.569030176525
$,
$\sigma=   0.545346  %%0.545346071318504
$,
and  $\mu= 1.13031   %%1.130312559801595
$.
(Middle row)
Derived functions $w\,F(w)$,
$l(w)\equiv \sqrt{w}\,\left[1-1/\sigma^{2}(w)\right]$,
and $\sigma(w)\,\mu(w)$. For the functions shown,
the value of $|l(w)|$ is less than $10^{-6}$ over the
interval $w\in [10^3,\,10^7]$.
(Bottom row) Dimensionless Ricci curvature scalar $R(w)$,
dimensionless Kretschmann curvature  scalar $K(w)$,
and negative of the
00-component of the dimensionless Einstein tensor $E^{\mu}_{\;\;\nu}(w)$
$\equiv$ $R^{\mu}_{\;\;\nu}(w) -(1/2)\,R(w)\,\delta^{\mu}_{\;\;\nu}$.
See Appendix~B of Ref.~\citen{Klinkhamer2014-prd}
for the explicit expressions of $R(w)$, $K(w)$, and $E^{0}_{\;\;0}(w)$.}
\label{fig02}
\vspace*{0mm}
%\end{figure*}
\vspace*{10mm}
%\begin{figure*}[hb]
\vspace*{0mm}
\begin{center} %%\end{center}
\includegraphics[width=0.925\textwidth]{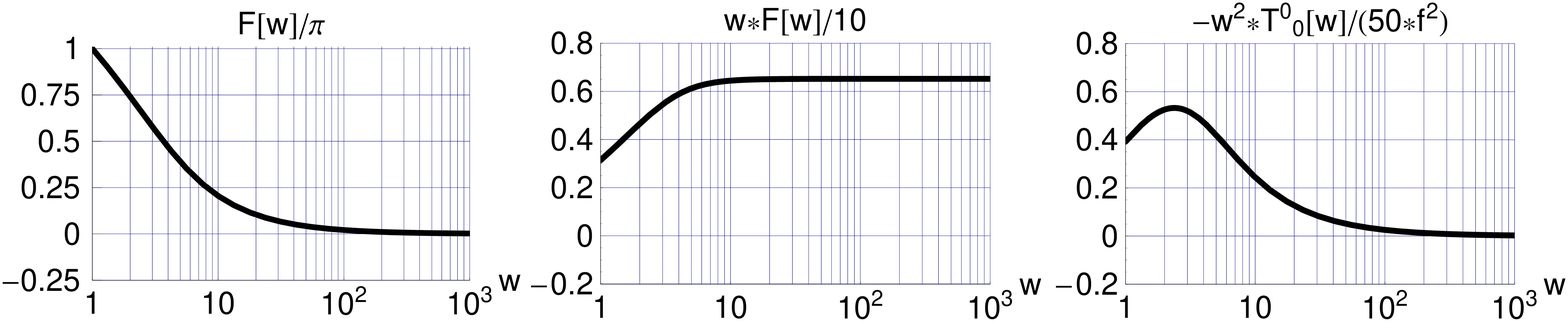}
\end{center}
\vspace*{-0mm}
\caption{Same as Fig.~\ref{fig02}, but with $\widetilde{\eta} =0$.
The ODEs for the metric functions $\sigma(w)$ and $\mu(w)$
can then be solved analytically and take the
Schwarzschild form \eqref{eq:Schwarzschild-type-solutions-mu-sigma}.
Here, the Schwarzschild parameter $\widehat{l}$
is taken to vanish exactly, which implies that
the corresponding ADM mass \eqref{eq:M-ADM} also vanishes exactly
[the spacetime manifold is flat with, in particular,
$R(w)=K(w)=E^{0}_{\;\;0}(w)=0$].
The corresponding ODE \eqref{eq:F-ODE-zero-etatilde} for $F(w)$,
with $\widetilde{m}=0$ and $\overline{\sigma}^2(w)=1$,
has the following
boundary conditions at the defect surface $w=y_0^2=1$:
$F=\pi$ and  $F^\prime=-1077934703/10^{9}$.
The right panel shows the energy density of the matter fields,
with the energy-momentum-tensor component $T^{0}_{\,\,0}(w)=-\rho(w)$
given by \eqref{eq:T-up0-down0}.
}
\label{fig03}
\vspace*{0mm}
\end{figure*}

Numerical results for Skyrmion spacetime defects
have been presented in Ref.~\citen{KlinkhamerQueiruga2018-prd}
(further details on the two-dimensional solution space
appear in Ref.~\citen{Guenther2017}).
In fact, Figs.~5 and 6 of Ref.~\citen{KlinkhamerQueiruga2018-prd} give
two numerical solutions with, respectively, negative and
positive asymptotic gravitational mass. These two solutions both
have boundary conditions \eqref{eq:bcs-F},
but differ in the boundary condition value of the
metric function $\sigma(w)$:
$\sigma(b^2)$ is smaller for the solution
of Fig.~5 than for the solution of Fig.~6.
For an appropriate intermediate value of $\sigma(b^2)$,
we expect to find a solution with vanishing
gravitational mass.
The \emph{Ansatz} functions for this numerical solution are shown
in Fig.~\ref{fig02}.
Practically, these functions have been obtained by integrating
inwards from boundary conditions at a large value $w_\text{max}$
[specifically, $F(w_\text{max})$ and $F'(w_\text{max})$ 
from \eqref{eq:F-asymptotic} at $w=w_\text{max}$
and $\sigma(w_\text{max})$ from \eqref{eq:sigma-asymptotic}
at $w=w_\text{max}$ with $\widetilde{l}=0$]
and tuning the asymptotic parameter $\widetilde{k}$
to get $F(1)=\pi$.

As mentioned in the caption of Fig.~\ref{fig02}, the
obtained value for $l_\infty$ is very small but not exactly
zero. This can, however, be achieved by
considering the special theory with $\widetilde{\eta}=0$.
Then, it is possible to 
set $\widehat{l}=0$ in the analytic solutions  
\eqref{eq:Schwarzschild-type-solutions-mu-sigma}
and also in the corresponding
ODE \eqref{eq:F-ODE-zero-etatilde} for $F(w)$.
The ODE \eqref{eq:F-ODE-zero-etatilde}
with $\widetilde{m}=0$ and $\overline{\sigma}^2(w)=1$
can be solved numerically and the result is
shown in Fig.~\ref{fig03}.

\begin{figure*}[ht]  %%[h!]
\vspace*{-0mm}
\begin{center} %%\end{center}
\includegraphics[width=0.925\textwidth]{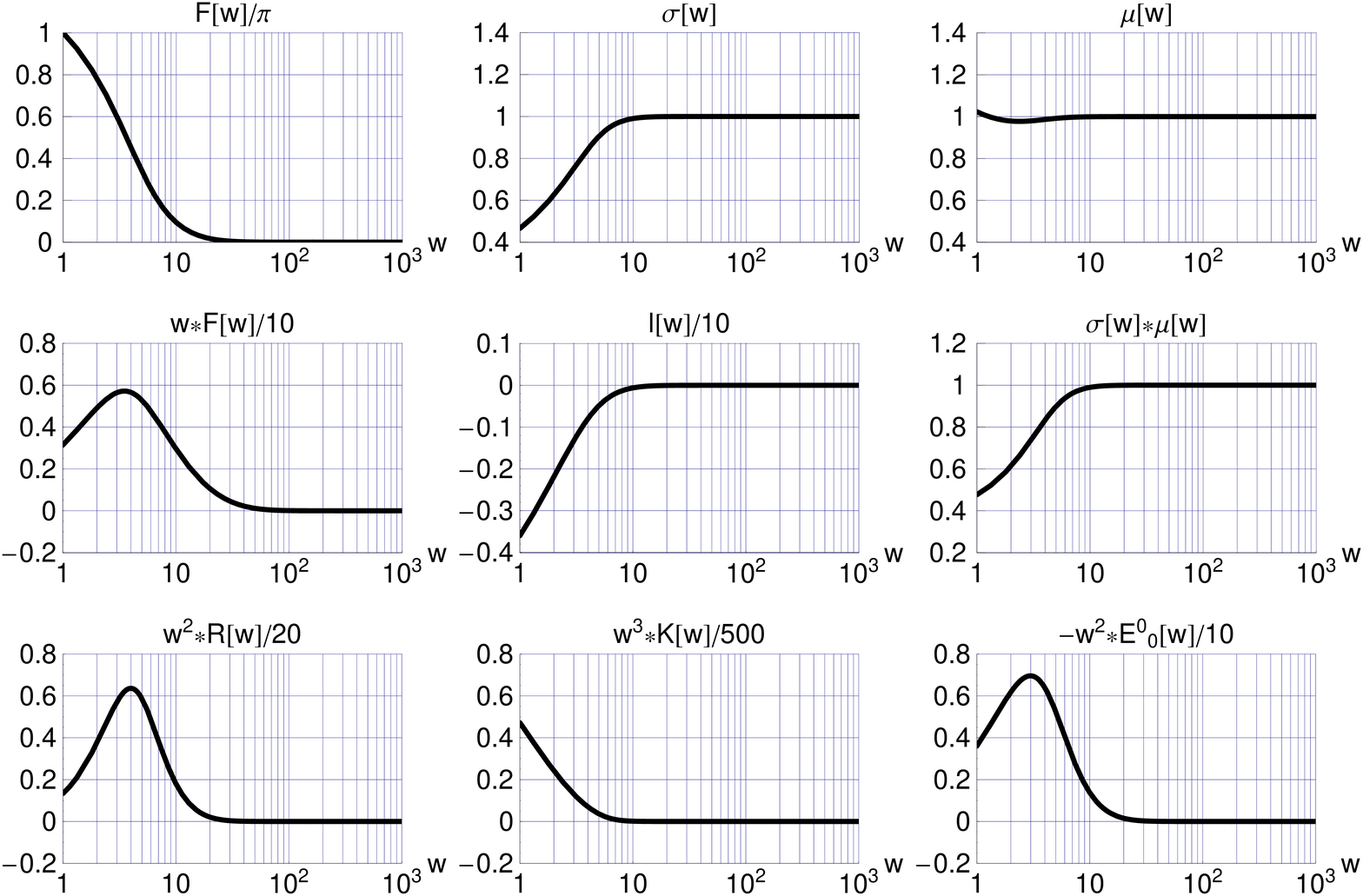}
\end{center}
\vspace*{-0mm}
\caption{Same as Fig.~\ref{fig02}, but with
$\widetilde{m}=1$. The boundary conditions at the defect surface $w=y_0^2=1$ are:
$F/\pi=     1.00000
$,
$F^\prime=  -0.752388
$,
$\sigma=    0.466343
$,
and  $\mu=  1.02282
$.
For the functions shown, the value of $|l(10^3)|$ is less than
$10^{-11}$.
}
\label{fig04}
\vspace*{0mm}
%\end{figure*}
\vspace*{5mm}
%\begin{figure*}[hb]
\vspace*{0mm}
\begin{center} %%\end{center}
\includegraphics[width=0.925\textwidth]{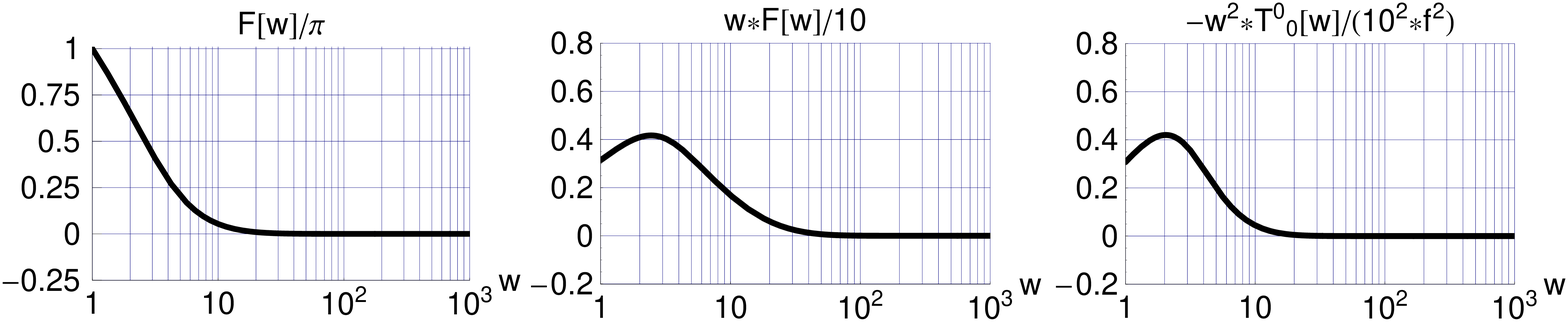}
\end{center}
\vspace*{0mm}
\caption{Same as Fig.~\ref{fig03}, but with
$\widetilde{m}=1$. The boundary conditions at the defect surface
$w=y_0^2=1$ are:
$F/\pi=     1.00000   %%1
$
and
$F^\prime= -1.43972$.
As in Fig.~\ref{fig03}, the ADM mass vanishes exactly and spacetime is flat.
}
\label{fig05}
\vspace*{0mm}
\end{figure*}

With a nonzero ``pion'' mass $m$, the matter fields
of the defect solutions have
an exponential tail, as shown by
\eqref{eq:F-asymptotic} for $\widetilde{m}\ne 0$.
Numerical results are given in Figs.~\ref{fig04} and
\ref{fig05}.
The numerical solutions of
Figs.~\ref{fig04} and \ref{fig05}
have an exponentially-vanishing
energy density of the matter fields,
together with a vanishing ADM mass.
In fact, the ADM mass of Fig.~\ref{fig05} is exactly zero.

%%%%%%%%%%%\newpage

The solution of Fig.~\ref{fig05} is a prime example of what we propose
to call a ``stealth defect.''   Outside this stealth defect,
the matter fields $\pi^a$ from \eqref{eq:Omega-pi-a} rapidly vanish 
and the spacetime is flat (Minkowskian away from the defect surface). 
Light rays follows straight paths and pass through the defect:
for radial paths crossing the defect surface,
the rays go straight through~\cite{Klinkhamer2014-mpla},
but, for non-radial paths crossing the defect surface,
the rays are parallel lines shifted by the
identification of antipodal points on the defect surface.

%%\newpage%%tmp
\section{Discussion}
\label{sec:Discussion}

In this article, we have discussed a particular type of
Skyrmion spacetime-defect solution~\cite{Klinkhamer2014-prd},
which has a vanishing asymptotic gravitational mass. As to the
origin of this zero mass, we observe that there is a
positive mass contribution from the matter fields
but a negative ``effective mass'' from the
gravitational fields at the defect surface $Y=0$.
As explained in our previous article~\cite{KlinkhamerQueiruga2018-prd}
and recalled in the third paragraph of Sec.~\ref{sec:Introduction},
this negative ``effective mass''
at the defect surface is needed in order to have
a globally regular solution if the defect scale $b$
is small enough.

For the solution of Fig.~\ref{fig05} in particular,
the matter energy density is
exponentially vanishing towards infinity (as long as $m^2 \ne 0$)
and the monopole term of the gravitational field is absent
($M_\text{ADM}=0$ exactly).
With a spherically symmetric defect surface and with a
spherically symmetric matter distribution, we also expect
no higher-order terms of the gravitational field
(quadrupole, octupole, etc.), but this expectation
needs to be confirmed.

Now, assume that all matter fields have some form of
non-gravitational interaction with each other.
If so, there will, in principle, be some interaction
between the ``pions'' of the theory considered in
\eqref{eq:action-Lgrav-Lmat-omegamu} and the elementary
particles of the standard model.
Then, consider what happens with a head-on collision
of a stealth defect and an observer made of
standard-model particles
(mostly up and down quarks, gluons, and electrons).
In close approximation, the observer will have no idea
of what is going to happen, until he/she is within
a distance of order $h/(m c)$ from the defect, where
$m$ is the ``pion'' mass scale from the
matter action \eqref{eq:Lmat}.
What happens during the collision itself
and afterwards depends on the details of the setup,
for example, the size of the observer compared to the
defect scale $b$. The only point we are sure of is
that, assuming the existence of this particular type of
spacetime-defect solution without long-range fields,
an observer has no advance warning if he/she approaches
such a stealth-type defect solution
(displacement effects of background stars are negligible,
at least initially).

%%\newpage%%tmp
\section{Outlook}
\label{sec:Outlook}

The topic of the present article, a stealth-type soliton solution,
is related to the larger physics issue of invisibility.
The prime example is an invisibility cloak which can be wrapped around 
a massive object in order to, more or less, hide the object from an outside
observer (see, e.g.,
Refs.~\citen{Pendry-etal2006,Leonhardt2006,Halimeh-etal2016,Schittny-etal2016} and references therein).
Still, the wrapped object has a gravitational mass and is, as such,
detectible by its gravitational force on a distant test particle.

In a way,
these two types of systems, cloaked object and stealth defect,
are orthogonal to each other.
The cloaked object, on the one hand, is more or less invisible for light
but can still be detected by its asymptotic gravitational mass,
$M_\text{ADM}>0$.
The stealth defect, on the other hand,
cannot be detected by its asymptotic  gravitational mass,
$M_\text{ADM}=0$, but is not really invisible for light
(assuming that the matter fields of the defect have interactions with
photons).

For the stealth defect,
the role of the ``gravitational invisibility cloak'' is played
by the defect surface and its nontrivial gravitational fields
(as discussed
in the second and third paragraphs of Sec.~\ref{sec:Introduction}):
these gravitational fields make
the matter fields just outside the defect surface
gravitationally invisible at infinity.
For the stealth-defect system,
this ``gravitational invisibility cloak'' is buried
inside the matter, instead of wrapping around the energy density
distribution of the matter.

As a final speculative comment, we could try to hide
a massive object by placing an appropriate spacetime defect next to it
and wrapping the whole in an invisibility cloak.
The defect must have the appropriate negative gravitational mass,
so that the total gravitational mass of object+defect+cloak is zero.
In addition, the cloak must be strong enough, because the material object
receives a repulsive gravitational force from the defect.

Expanding on the last comment, we may consider the present
article as a small step towards the  ``applied physics of the future,''
which deals with designing and modelling
spacetime in addition to designing and modelling ponderable
matter. In this respect, we should mention that
the crucial open question is the
origin and role of nontrivial spacetime topology
(see, in particular, Chap.~6 of Ref.~\citen{Visser1995}).
Specialized to our Skyrmion spacetime-defect solution, the questions
are what sets the constant defect scale $b$ and
can this defect scale become a dynamic variable?
These and other theoretical questions need to be addressed before
we can really start thinking about
the ``applied physics of the future.''

\vspace*{-0mm}
\section*{Acknowledgment}
\vspace*{-0mm}
\noindent
FRK thanks M. Wegener for several discussions on invisibility cloaks
over the last years.

%%\newpage%%tmp
%\vspace*{-0mm}


\begin{thebibliography}{99}


\bibitem{Klinkhamer2014-prd}
F.R.~Klinkhamer,
``Skyrmion spacetime defect,''
\textit{Phys.\ Rev.\ D} \textbf{90}, 024007 (2014),
arXiv:1402.7048.  %% [gr-qc]].
%%CITATION = doi:10.1103/PhysRevD.90.024007;%%

\bibitem{KlinkhamerQueiruga2018-prd}
F.R.~Klinkhamer and J.M.~Queiruga,
``Antigravity from a spacetime defect,''
\textit{Phys. Rev. D} \textbf{97}, 124047 (2018), 
arXiv:1803.09736.  %% [gr-qc].
%%CITATION = ARXIV:1803.09736;%%

\bibitem{Guenther2017}
M. Guenther,
``Skyrmion spacetime defect, degenerate metric,
and negative gravitational mass,''
Master Thesis, KIT, September 2017;
available from
\verb"https://www.itp.kit.edu/en/"
\verb"publications/diploma"

\bibitem{Klinkhamer2014-mpla}
F.R.~Klinkhamer,
``A new type of nonsingular black-hole solution in general relativity,''
\textit{Mod.\ Phys.\ Lett.\ A} \textbf{29}, 1430018 (2014),
arXiv:1309.7011.  %% [gr-qc]].
%%CITATION = doi:10.1142/S0217732314300183;%%

\bibitem{DeWitSmith1986}
B.~De Wit and J.~Smith,
\emph{Field Theory in Particle Physics, Volume 1}
(North-Holland Physics Publishing,  
Amsterdam, The Netherlands, 1986).
%%CITATION = INSPIRE-237412;%%


\bibitem{ADM1959}
R. Arnowitt, S. Deser, and C.W. Misner,
``Dynamical structure and definition of energy in general relativity,''
\textit{Phys. Rev.} \textbf{116}, 1322 (1959).
 %CITATION = doi:10.1103/PhysRev.116.1322;%%



\bibitem{Pendry-etal2006}
J. B. Pendry, D. Schurig, and D. R. Smith,
``Controlling electromagnetic fields,''
\textit{Science} \textbf{312}, 1780 (2006).

\bibitem{Leonhardt2006}
U. Leonhardt,
``Optical conformal mapping,''
\textit{Science} \textbf{312}, 1777 (2006),
arXiv:
physics/0602092v1.

\bibitem{Halimeh-etal2016}
J.C. Halimeh, R.T. Thompson, and M. Wegener,
``Invisibility cloaks in relativistic motion,''
\textit{Phys. Rev. A} \textbf{93}, 013850 (2016),
arXiv:1510.06144.

\bibitem{Schittny-etal2016}
R. Schittny, A. Niemeyer, F. Mayer, A. Naber, M. Kadic, and M. Wegener,
``Invisibility cloaking in light-scattering media,''
\textit{Laser Photonics Rev.} \textbf{10}, 382 (2016).


\bibitem{Visser1995}
M.~Visser,
\emph{Lorentzian Wormholes: From Einstein to Hawking}
(Springer-Verlag, New York, USA, 1995).

\end{thebibliography}
\end{document}